\documentclass[aps,pre,singlecolumn,superscriptaddress]{revtex4}
 
\usepackage{graphicx} 
\usepackage{physics, color, float}

\newcommand{\ds}{\displaystyle}
\newcommand{\beq}[0]{\begin{equation}}
\newcommand{\eeq}[0]{\end{equation}}

\begin{document}

\title{Hydromagnetic shock waves in a cold weakly collisional
plasma}

\author{M. Calabrese}
\affiliation{Department of Mathematics and Statistics, University of Massachusetts, Amherst MA 1003-4515, USA}
 \author{V. Koukouloyannis}
\affiliation{Department of Mathematics, University of the Aegean, Karlovasi, 83200 Samos, Greece}
\affiliation{Department of Mathematics and Statistics, University of Massachusetts, Amherst MA 1003-4515, USA}
\author{S. S. Abbas}
\affiliation{Department of Physics, G. C. University, Katchery Road, Lahore 54000}
\author{G.\ Abbas}
\affiliation{Department of Physics, G. C. University, Katchery Road, Lahore 54000}
\author{P.G.\ Kevrekidis}
\affiliation{Department of Mathematics and Statistics, University of Massachusetts, Amherst MA 1003-4515, USA}

\begin{abstract}
In this work we revisit the topic of existence of hydrodynamic shock waves in a cold weakly collisional plasma. For this purpose we consider the well established Adlam-Allen model 
with the addition of a dashpot term associated with the dissipation of the motion of
the electrons relative to the ions. We establish the connection between this model and the Korteweg-de-Vries Burgers equation via an asymptotic multiscale analysis. This fact suggests the possibility that this system may support
shock wave solutions. Accordingly, by considering a corresponding dynamical system arising through a co-traveling frame reduction,
we identify such orbits via a phase-plane  analysis.
We then leverage such initial conditions within systematic
simulations of the original modified Adlam-Allen model,  revealing a variety of supported robust wavefronts depending on the magnitude of the dissipation considered.
\end{abstract}

\maketitle

\section{Introduction}

The original proposal of Fermi-Pasta and Ulam and associated
numerical work by Tsingou~\cite{fermi1955studies} is widely
credited as the birth point of Nonlinear Science, as we 
know it today. A pivotal point in the relevant history
is widely known to be the work of Zabusky and Kruskal 
in 1965 which led to the emergence of the theory of
solitary waves (known as solitons in 
integrable systems)~\cite{zabusky1965interaction}.
While it is clear that the latter work was inspired
by research efforts in collisionless-plasma magnetohydrodynamics (through the citation of an 
unpublished report by Gardner and Morikawa therein),
what appears to be far less well known is that already
 7 years earlier 
than~\cite{zabusky1965interaction},
Adlam and Allen had proposed a remarkable  model
featuring solitary hydrodynamic waves in plasmas~\cite{Allen1958}. Interestingly the so-called
Adlam-Allen model was also missed in subsequent,
well-known work on the propagation of ion-acoustic solitary
traveling waves of small amplitude in plasmas by Washimi and 
Taniuti~\cite{washimi1966propagation}.




In recent years, there has been a revival of interest
in the Adlam-Allen (AA) model, such as, e.g., through the re-derivation of
its exact analytical solutions and the proposal of
corresponding
 applications to observations in di-polarization fronts  in
the Earth's inner magnetosphere in the work of~\cite{Abbas2020,Abbas2022}. Another vein of recent
efforts was towards considering connections via asymptotic
multiscale reductions  to the setting of the 
famous Korteweg-de Vries (KdV) equation~\cite{korteweg1895change},
as well as in the study of multiple solitary waves in the model
and their interaction in the work of~\cite{koukouloyannis2022interactions}.

In the present work our scope is to explore
an important variation to the above setting
that dates back to the formulation of
Tidman and Krall~\cite{Tidman1971}. In particular,
the idea is that in the direction transverse to the flow 
the particles (most notably the
much faster electrons, in comparison to the nearly
stationary ions) may suffer scattering in the
microturbulence range within the plasma, leading
to an effective weak dissipation effect. These authors
described such features through a collision kernel
and obtained a corresponding reduced description via
a dashpot type dissipation term within the
evolutionary dynamics of the electrons. Our principal
idea herein is that in the presence of such dissipation,
the KdV analysis of~\cite{Allen2020} can be suitably
generalized in the form of a KdV-Burgers (KdVB) equation
which, in turn, leads to oscillatory shock waves in the
underdamped regime, while it leads to monotonic ones
in the overdamped case. Studies of these types of
subcritical and supercritical shock waves
have arisen in a variety of problems ranging
from electromagnetic waves in optical fibers
in the presence of stimulated Raman scattering~\cite{Kivshar1993} to
hot collisional plasmas in astrophysical applications~\cite{Treumann2009}, among numerous
others~\cite{Blandford1987,Krall1997,Dimmock2019,Mahdipoori2019,Kono2010}. 

We start our presentation by revisiting the relevant
collisional model and examining how the dissipative term
induces a modification of the resulting partial differential
equations for the (inverted) dimensionless density and
magnetic field. Then, in section III, we perform the
multiple scales reduction of the original Adlam-Allen with Damping (AAD)
model to the KdVB one. This enables us to then move
to the cotraveling frame of the system and identify the
conditions for the existence of a solitary wave. Not only
are we able, on the basis of phase-plane considerations,
to obtain the traveling shock wave solution; indeed, we  explore different scenarios pertaining to under-
and over-damping (and their potential physicality), as well
as the implications of the change of sign of the 
dissipative term (to an anti-damped one). All of our
findings for the reduced multiscale system are 
eventually brought to bear back to the original AAD
model in section IV, where the robustness of the corresponding structures
is corroborated through direct numerical simulations.
We believe that this constitutes a fundamental manifestation
of an extensive family of shock solutions within this problem
and as such may be of interest to a diverse audience of 
plasma physics and nonlinear science practitioners.



\section{Derivation of the model}
Our modeling starting point consists of the dynamical
equations for the electrons and the ions in the 
Newtonian form:
 \begin{equation} \label{eq1}
m_sD_s\textbf{v}_s=q_s(\textbf{E}+\textbf{v}_s\times \textbf{B}) +\frac{m_s}{N}\int Cv_sdv_s
 \end{equation}
 where $D_s=\frac{\partial}{\partial \tau}+\textbf{v}_s\cdot\nabla$, $\nabla=\left (\frac{\partial}{\partial \chi},0,0 \right )$, $\textbf{v}_s=(v_{s1},v_{s2},v_{s3})$ is the velocity of the particles with $s=e,i$ for electrons or ions respectively, $\textbf{E}=(E_1,E_2,0)$ and
 $\textbf{B}=(0,0,B_3)$ are the electric field and the magnetic field (with mass represented by $m$ and 
 charge by $q$), respectively, and $\frac{1}{N}\int Cv_sdv_s$ represents the collision term, as formulated in the
 classic book of~\cite{Tidman1971}. Now, using Ampere's and Faraday's laws  and taking into consideration the quasineutrality condition (i.e. $n_i\approx n_e \approx n$) one gets:
 \begin{align*}
    &v_{i1}\approx v_{e1}=:v_1\\
    &\frac{\partial B_3}{\partial \chi}=\mu_0en(v_{e2}-v_{i2}) \\
    &\frac{\partial E_2}{\partial \chi}=-\frac{\partial B_3}{\partial \tau}
 \end{align*}
where $\mu_0$ stands for the magnetic constant and $e$
for the electron charge.
{The first equation belongs to the case when currents along x-direction are taken to be zero. When $J_x$ = 0, then $v_{i1}=v_{e1}$ due to the quasineutrality condition in line
also with the exposition of~\cite{Tidman1971}.
As is discussed therein, the electrons are assumed to gyrate
rapidly producing a large transverse drift curent
leading to
$v_{i2}=\mu v_{e2}$, where $\mu=\frac{m_e}{m_i}$. 
Assuming dissipation in this much faster $y$-motion
(since as~\cite{Tidman1971} points out
$v_{e2} \gg v_{i1} \approx v_{e1}$ and also
$v_{e2} \gg v_{i2}$), the
dissipation term associated with $v_{i2}$  will be 
negligible in comparison to that associated with $v_{e2}$}.
Now developing in coordinates equation (\ref{eq1}), one has the following equations (component wise):
\begin{align}
     &m_i\left (\frac{\partial v_{1}}{\partial \tau} + v_{1}\frac{\partial v_1}{\partial \chi}\right )=e(E_1+v_{i2}B_3) \label{eqi1} \\
      &m_e\left (\frac{\partial v_{1}}{\partial \tau} + v_{1}\frac{\partial v_{1}}{\partial \chi}\right )=-e(E_1+v_{e2}B_3) \label{eqe1}\\
     & m_i\left (\frac{\partial v_{i2}}{\partial \tau} + v_{1}\frac{\partial v_{i2}}{\partial \chi}\right )=e(E_2-v_{1}B_3) \label{eqi2} \\
     & m_e\left (\frac{\partial v_{e2}}{\partial \tau} + v_{1}\frac{\partial v_{e2}}{\partial \chi}\right )=-e(E_2-v_{1}B_3) -\nu m_e v_{e2} \label{eqe2}
 \end{align}
Observe that in the case of small-amplitude turbulence, the relative drift of particles $v_{ek}-v_{ik}$ for $k=1,2,3$  is considered significant in the second direction, while the other two components are neglected due to their small values. 
{The consideration of dissipation along the y-direction and its association with electrons, not ions in the context of an AAD shock wave propagating perpendicularly to the ambient magnetic field is motivated by the setting under consideration
herein, as well as the distinctions in dynamics between electrons and ions.  
More specifically, dissipation may occur due to mechanisms like collisional viscosity, thermal conductivity, or wave-particle interactions. These dissipative processes may differ in effectiveness along and perpendicular to the field lines. Dissipation effects along the y-direction are often associated with transverse dynamics and micro-instabilities. These y-components are essential for accounting for energy loss and maintaining shock structure stability in realistic scenarios. Dissipation along the longitudinal  direction is minimal in many treatments because longitudinal dynamics are dominated by bulk motion with relatively weak energy gradients in the field-aligned direction. The associated dissipation term in the present wave geometry is supposed to depend chiefly on transverse currents, making longitudinal dissipative mechanisms secondary.}
Assuming  weak collisions, this phenomenon leads
to small dissipation and, as argued in~\cite{Tidman1971}, the collision term $\frac{1}{N}\int Cv_sdv_s$ can be approximated as \(- \nu v_{e2}\), where \(\nu\) represents the effective collisional frequency associated with electrons, as described in equation (1.16) of Tidman and Krall (1971) \cite{Tidman1971}. \newline \newline
Dividing both (\ref{eqi2}) and (\ref{eqe2}) by $m_i$ and adding
them together, we infer that: 
\begin{equation} \label{eq2}
\left (\frac{\partial v_{i2}}{\partial \tau}+ v_{1}\frac{\partial v_{i2}}{\partial \chi}\right )=-\frac{m_e}{m_i}\left (\frac{\partial v_{e2}}{\partial \tau} + v_{1}\frac{\partial v_{e2}}{\partial \chi}\right )-\frac{m_e}{m_i} \nu v_{e2} \end{equation}
If we now add Eqs.~(\ref{eqi1})-(\ref{eqe1}), as well 
as divide (\ref{eqi2}) and (\ref{eqe2}) by their respective
masses and then subtract them side by side, we obtain the following
system:
\begin{align} 
    &\left (\frac{\partial v_{1}}{\partial \tau} + v_{1}\frac{\partial v_{1}}{\partial \chi}\right ) =\frac{-v_2eB_3}{m_i+m_e} \label{eq3} \\
   &\left (\frac{\partial v_{2}}{\partial \tau} + v_{1}\frac{\partial v_{2}}{\partial \chi}\right )=-e\frac{m_i+m_e}{m_im_e}(E_2-v_1B_3)-\nu v_{e2}\label{eq4}
\end{align}
where $v_2=v_{e2}-v_{i2}$. So considering equations (\ref{eq3}) and (\ref{eq4}), the equations from Ampere's law and Faraday's law and the transport equation for the evolution of the density of the particles
lead to the following system of 5 equations:
\begin{align}
      &\frac{\partial v_{1}}{\partial \tau} + v_{1}\frac{\partial v_{1}}{\partial \chi} =\frac{-v_2eB_3}{m_i+m_e}  \label{eq 5} \\
     &\frac{\partial v_{2}}{\partial \tau} + v_{1}\frac{\partial v_{2}}{\partial \chi}=-e\frac{m_i+m_e}{m_im_e}(E_2-v_1B_3)-\nu v_{e2} \label{eq 6}\\
     &\frac{\partial E_2}{\partial \chi}=-\frac{\partial B_3}{\partial \tau}  \label{eq 7} \\
     &\frac{\partial B_3}{\partial \chi}=\mu_0env_2 \label{eq 8}\\
     &\frac{\partial n}{\partial \tau}=-\frac{\partial (nv_1)}{\partial \chi} \label{eq 9}
 \end{align}
 Now following the paper \cite{Allen1958} let us introduce a Lagrangian coordinate $\mathcal{X}$ that is constant for any particle and is equal to the $\chi$ coordinate of that particle at $\tau=0$, under these assumptions we have:
 
     $$ \left ( \frac{\partial}{\partial \mathcal{T}}\right )=\left ( \frac{\partial}{\partial \tau} \right )+\left ( \frac{\partial \chi}{\partial \tau} \right )\left ( \frac{\partial}{\partial \chi} \right )=\left ( \frac{\partial}{\partial \tau} \right )+v_1\left ( \frac{\partial}{\partial \chi} \right ) $$
     $$ \left ( \frac{\partial}{\partial \chi}\right )=\left ( \frac{\partial \mathcal{X}}{\partial \chi}\right )\left ( \frac{\partial}{\partial \mathcal{X}}\right )=\frac{n}{n_0}\left ( \frac{\partial}{\partial \mathcal{X}}\right ).$$
    {where $n_0$ is the initial density.}\newline 
     This, in principle, introduces a Lagrangian frame where,
     however, space is suitably rescaled by the particle density.
Leveraging this change of coordinates, we can write our set of equations (\ref{eq 5})-(\ref{eq 9}) as:
 \begin{align}
    &\frac{\partial v_1}{\partial \mathcal{T}} =\frac{-v_2eB_3}{m_i+m_e} \label{eq 10} \\
     &\frac{\partial v_2}{\partial \mathcal{T} }=-e\frac{m_i+m_e}{m_im_e}(E_2-v_1B_3)-\nu v_{e2}  \label{eq 11}\\
     &\frac{\partial E_2}{\partial \mathcal{X}}=-\frac{n_0}{n}\frac{\partial B_3}{\partial \mathcal{T}}+v_1\frac{\partial B_3}{\partial \mathcal{X}}  \label{eq 12}\\
     &\frac{\partial B_3}{\partial \mathcal{X}}=\mu_0en_0v_2  \label{eq 13} \\
     &\frac{\partial}{\partial \mathcal{T}}\left ( \frac{n_0}{n} \right )=\frac{\partial (v_1)}{\partial \mathcal{X}}  \label{eq 14}
 \end{align}
Observe that from equation $(\ref{eq2})$ we can obtain that $\frac{\partial v_2}{\partial \mathcal{X}}\approx\frac{\partial v_{e2}}{\partial \mathcal{X}}$ 
{and this allows us to neglect the term $\frac{\partial v_i}{\partial \mathcal{T}}$ in equation (\ref{eq 11})}.
\newline \newline 
From a physical point of view, for the second velocity components of ions and electrons in a weak collisional shock, the difference \(v_{e2} - v_{i2} = v_{e2}\) is due to the negligible contribution of \(v_{i2}\).
In recent works \cite{Abbas2022,Allen2020,Abbas2020}, even without collisions, detailed mathematical treatment showed that \(v_{i2} = - \mu v_{e2} = -\frac{v_{e2}}{1836}\), reflecting the mass difference between electrons and ions. 
Given the assumption of weak collisions, electrons—being much lighter than ions—are more affected by collisions. Due to their higher mobility, electrons acquire much higher velocities than ions, which can be considered nearly stationary by comparison and less responsive in shock dynamics. Electrons, having a faster response to electric fields because of their lower mass, justify neglecting the ion velocity component \(v_{i2}\). Ions, due to their greater mass and slower response to collisions, decelerate more rapidly,
justifying the assumption:
\[
v_2=v_{e2} - v_{i2} \approx v_{e2}
\]
Additionally, the electron-ion collision frequency is substantially lower than electron-electron collisions, allowing electrons to be more dynamically influenced by the electric field, while momentum transfer between electrons and ions remains minimal~\cite{Tidman1971}.
As a result, in weakly collisional environments, the electrons' response dominates the system's behavior, while ions remain largely unaffected over short timescales.
\newline \newline
Now we  introduce the following dimensionless quantities:
 $d=\frac{v_A}{\Omega_e\sqrt{\mu}}$ called the Adlam-Allen distance, $v^*=\frac{v_A}{\sqrt{\mu}}$, $v_A=\frac{B_0}{\sqrt{\mu_0n_0(m_e+m_i)}}$ being the Alfvén velocity, while $\Omega_e=\frac{eB_0}{m_e}$ is defined as the electron angular frequency. 
 \newline
Under these parameters the equations (\ref{eq 10})-(\ref{eq 14}) in dimensionless form are:
\begin{align}
&\frac{\partial E}{\partial x} =+v_{1}^{^{\prime}}\frac{\partial B}{\partial x}-R\frac{\partial B}{\partial t} \label{eq 15}\\
&\frac{\partial R}{\partial t}
=\frac{\partial v_{1}^{^{\prime}}}{\partial x}  \label{eq 16}\\
&\frac{\partial v_{1}^{^{\prime}}}{\partial t}  =-\frac{v_2^{^{\prime}}%
B}{\left(  1+\mu\right)  }  \label{eq 17}\\
&\frac{\partial B}{\partial x}  =\frac
{v_2^{^{\prime}}}{\left(  1+\mu\right)  } \label{eq 18}\\
&\frac{\partial v_2^{^{\prime}}}{\partial t}   =\left(  1+\mu\right)
(-E+v_{1}^{^{\prime}})-\frac{\nu}{\sqrt{\mu}
\Omega_e}v_{e2}^{^{\prime}}  \label{eq 19}%
\end{align}
where $R=\frac{n_0}{n},x=\frac{\mathcal{X}}{v_A/\Omega_e\sqrt{\mu}}$, ${v_1}^{^{\prime}}=\frac{v_1}{v_A}$, ${v_{2,3}}^{^{\prime}}=\frac{v_{_{2,3}}}{v_{A}/\sqrt{\mu}}$, $B=\frac{B_3}{B_0}$,$t=\frac{\mathcal{T}}{\frac{\sqrt{m_em_i}}{eB_0}}$, $E=\frac{E_2}{v_AB_0}.$\newline \newline
Starting from (\ref{eq 15})-(\ref{eq 19}) we can derive the Adlam-Allen equations with a dissipative collisional term, as follows.
Computing a time derivative in equation $(\ref{eq 16})$ and then substituting equations (\ref{eq 17}) and (\ref{eq 18}) one obtains:
\begin{equation}
\frac{\partial^{2}R}{\partial t^{2}}
=-\frac{\partial}{\partial x}\left(B
\frac{\partial B}{\partial x}\right) \label{eq 20}
\end{equation}
that is the first equation of Adlam-Allen model \cite{Allen1958}. 
For the second equation, starting from equation (\ref{eq 18}) and computing a time derivative and a space derivative with the appropriate substitution of Eq.~(\ref{eq 19}) one gets:
\begin{equation}
\frac{\partial^{3}B}{\partial x^{2}\partial t}
=\frac{\partial}{\partial t}\left(BR\right)-\frac{\nu}{\sqrt{\mu}
\Omega_e\left(  1+\mu\right)  }\frac{\partial
v_{e2}^{^{\prime}}}{\partial x}. \label{eq 21}
\end{equation}
Then, using $(\ref{eq2})$ and 
{differentiating in $x$ equation} $(\ref{eq 18})$ results in:
\begin{equation}
    \frac{\partial^{3}B}{\partial x^{2}\partial t}=\frac{\partial}{\partial t}\left( BR\right)-\frac{\nu}{\sqrt{\mu}
\Omega_e  } \frac{\partial^2 B}{\partial x^2}. \label{eq 22} \end{equation}
Equation ($\ref{eq 22}$) is exactly the second equation of AA model with a collision term, which is our central model of interest herein
towards the description of cold, weakly collisional plasmas 
and their potential shock waves.

\section{Connection to KdVB equation}
In the previous section we found the equations for the AA model with the dissipation term given by:
\begin{align}
 &\frac{\partial^{2}R}{\partial t^{2}}
=-\frac{\partial}{\partial x}\left(  B%
\frac{\partial B}{\partial x}\right) \label{eq 23} \\
  &\frac{\partial^{3}B}{\partial x^{2}\partial t}=\frac{\partial}{\partial t}\left(BR\right)-k\frac{\partial^2 B}{\partial x^2} \label{eq 24} 
\end{align}
where $k=\frac{\nu}{\sqrt{\mu}
\Omega_e  } $. \newline
This model admits a simple homogeneous solution:
$$R=R_0 \ , \ B=B_0$$
where $R_0$ and $B_0$ are such that $R\longrightarrow R_0$ and $B\longrightarrow B_0$  and $ x\longrightarrow \pm \infty.$ Now it is convenient to translate the variables in the following way: $R(x,t)=u(x,t)+R_0$ and $B(x,t)=w(x,t)+B_0$ where $u$ and $w$ satisfy vanishing boundary conditions. Under this change of variables, equations $(\ref{eq 23})$ and  $(\ref{eq 24})$ become:
\begin{align}
 &u_{tt}=-\left ( \frac{w^2}{2}+B_0w \right )_{xx} \label{eq 25}\\
&w_{xxt}=(B_0u+R_0w+uw)_{t}-kw_{xx}.\label{eq 26}
\end{align}
Following similar considerations as in \cite{Allen2020} we want to establish an asymptotic connection of equations (\ref{eq 25}) and (\ref{eq 26}) with the KdVB equation, as a toolbox towards
understanding the traveling wave coherent structures
within the model. To do so, we judiciously introduce the slow variables:
\begin{equation}
    X=\epsilon^{\frac{1}{2}}(x-Ct), \ \ T=\epsilon^{\frac{3}{2}}t, \ \ K=\epsilon^{\frac{1}{2}}k \label{eq 27}
\end{equation}
where $C^2=\frac{B_0^2}{R_0}.$  \newline \newline 
With a time differentiation of $(\ref{eq 26})$ and using $(\ref{eq 27})$, equations $(\ref{eq 25})$ and $(\ref{eq 26})$ can be written as:
\begin{align*}
    \epsilon^2C^2w_{XXXX}+\epsilon^4w_{TTXX}-2\epsilon^3Cw_{XXXT}=&-\epsilon B_0\left ( \frac{w^2}{2}\right )_{XX}+R_0(\epsilon^3w_{TT}-2C\epsilon^{2}w_{XT})\\
    &-\epsilon w \left ( \frac{w^2}{2}+B_0w \right)_{XX}+2\epsilon C^2 u_Xw_X-2\epsilon^2Cu_Xw_T \\
    &-2\epsilon^2Cu_Tw_X+2\epsilon^3u_Tw_T\\ 
    &+u(\epsilon C^2w_{XX}-2C\epsilon^2 w_{XT}+\epsilon^3 w_{TT}) \\
    &-K(\epsilon^{2}Cw_{XXX}+\epsilon^{3}w_{TXX}).
    \end{align*}
Now, expanding $w$ and $u$ as $\displaystyle w=\sum_{i=1}^\infty w_i\epsilon^i$ and $\displaystyle u=\sum_{i=1}^\infty u_i\epsilon^i$ and using from \cite{Allen2020} that $u_1=-\frac{R_0}{B_0}w_1$, and collecting the terms of order $\epsilon^3$ one has:
\begin{align}
2Cw_{1XT}+C^2w_{1XXXX}+3B_0(w_1w_{1X})_X+KCw_{1XXX}=0. \label{eq32}
\end{align}
Thus, after an integration with respect to $X$, equation $(\ref{eq32})$ leads to the following KdVB equation:
\begin{equation}
 2Cw_{1T}+C^2w_{1XXX}+3B_0w_1w_{1X}+KCw_{1XX}=0.    
\end{equation}
This, in turn, strongly suggests that the principal solutions
of the KdVB model, such as the oscillatory shock structures
(in the underdamped regime) and the monotonic shock waves
(in the overdamped case), as detailed in textbooks such as~\cite{Carretero-Gonzalez2024}, should also be present herein.
Indeed, we now turn to numerical considerations that allow
the establishment of these different classes of solutions
of the weakly collisional Adlam-Allen model.

\section{Numerical Computations}
\subsection{The dynamical system treatment - Left-moving waves}
\begin{figure}[!h]
\begin{center}
\includegraphics[scale=0.35]{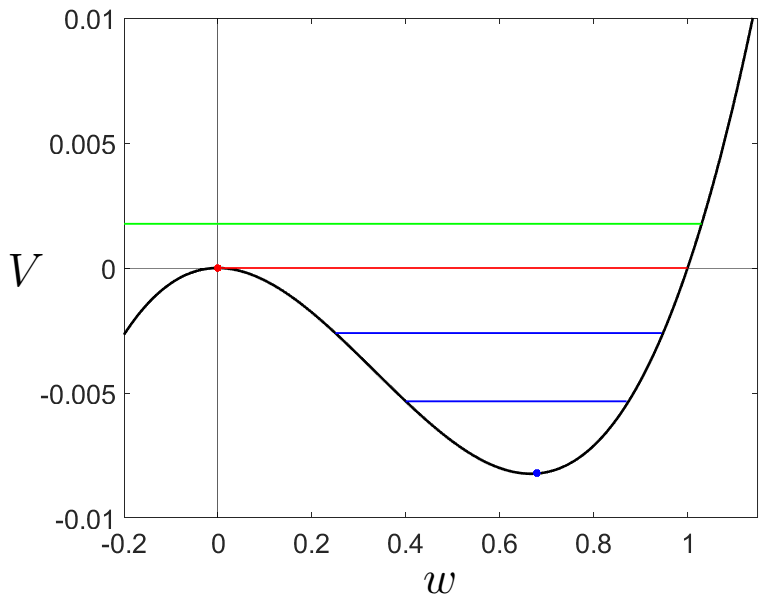}\hspace{0.5cm} \includegraphics[scale=0.35]{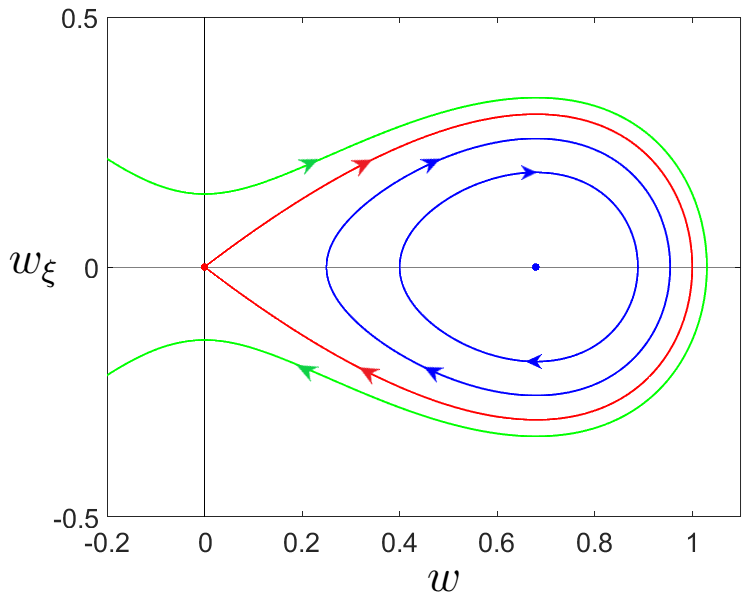}
\caption{Left panel: The potential function of the corresponding conservative ($k=0$) system with some characteristic energy levels. The colors correspond to the relative orbits in the phase-plane of the system. Right panel: The orbits of positive energy are irrelevant since they are not confined to positive values of $w$. The zero energy corresponds to a homoclinic orbit. Finally, the negative energies correspond to closed periodic orbits.}
\label{fig:conservative}
\end{center}
\end{figure}

Returning to the system \eqref{eq 25}-\eqref{eq 26}, we consider a 
co-traveling frame of reference moving with velocity $c>0$ equal to the constant velocity of the wave  moving to the left, by using the transformation $u(x,t)\mapsto u(x+ct)\equiv u(\xi)$ and $w(x,t)\mapsto w(x+ct)\equiv w(\xi)$ . Then, the differential operators transform as $\ds\pdv{}{x}\rightarrow \pdv{}{\xi}$ and $\ds\pdv{}{t}\rightarrow c\pdv{}{\xi}$ and \eqref{eq 25}-\eqref{eq 26} become
\begin{eqnarray}
c^2u_{\xi\xi}&=&-\left(B_0w+\frac{w^2}{2}\right)_{\xi\xi}\label{ds1}\\
w_{\xi\xi\xi}&=&\left(B_0u+R_0w+uw\right)_\xi-\frac{k}{c}w_{\xi\xi}.\label{ds2}
\end{eqnarray}
By integrating \eqref{ds1} twice and \eqref{ds2} once, 
implicitly assuming vanishing boundary conditions, given our
interest in states that asymptote to $R_0$ and $B_0$ for the 
two fields,
we get 
\begin{eqnarray}
u&=&-\frac{1}{c^2}\left(B_0w+\frac{w^2}{2}\right)\label{ds12}\\
w_{\xi\xi}&=&\left(B_0u+R_0w+uw\right)-\frac{k}{c}w_{\xi}.\label{ds22}
\end{eqnarray}
Substituting \eqref{ds12} into \eqref{ds22}, we acquire finally 
\beq
w_{\xi\xi}=\frac{B_0^2(c^2-C^2)}{c^2C^2}w-\frac{3B_0}{2c^2}w^2-\frac{1}{2c^2}w^3-\frac{k}{c} w_{\xi}, \label{ODE}
\eeq
which can be considered as a dissipative dynamical system with respect to $w$, while $\xi$ is playing the effective role of an (independent) evolution variable. The corresponding conservative (for $k=0$) system can naturally be considered as a mechanical one with potential $V(w)$ 
\beq
V=-\frac{B_0^2(c^2-C^2)}{2c^2C^2})w^2+\frac{B_0}{2c^2}w^3+\frac{1}{8c^2}w^4,
\eeq
in line with the earlier considerations of~\cite{Allen2020}.
In the physically meaningful interval of $c$, which is $1<c<2$ as in \cite{Allen2020}, this potential posseses a local maximum and two local minima, while $\lim_{w\pm\infty}V(w)=+\infty$. In addition, the only relevant values of $w$ are the ones with $w\geq0$, given the
definitions of $u$ and $w$. Under these considerations, the critical points that are relevant to our study are: the local maximum which lies at $w_{max}=0$ and the local minimum at $w_{min}=\frac{-3B_0C+\sqrt{8B_0c^2+(9B_0-8)B_0C^2}}{2C}>0$. For the rest of our study we will consider $B_0=R_0=1\Rightarrow C=1$ and $c=1.5$ (with the
latter being a mere representative parameter value). For these parameter values we get $w_{min}=0.67945$ (see Fig.~\ref{fig:conservative}). 

In the left panel of Fig.\ref{fig:conservative} the potential, for the specific (conservative) parameter choice, is shown together with some characteristic energy levels, while in the right panel of the figure the corresponding phase-plane orbits are depicted. The orbits (in green) which have energy greater than zero are not relevant to our study since they possess a part which lies in the unphysical area of negative values of $w<0$. The zero energy orbit (in red) is a homoclinic orbit starting from and returning to the unstable (saddle) equilibrium $w_{max}$ and corresponds to a solitonic solution, the celebrated solitary
wave solution of the Adlam-Allen model~\cite{Allen1958,adlam1960,Allen2020}. Finally, the negative energy orbits correspond to closed (periodic) orbits, which provide cnoidal wave solutions. 

When the dissipation coefficient becomes positive, i.e., $k>0$, the situation changes dramatically. 
In what follows, we will consider corresponding
cases that are underdamped (such as $k=0.03$
and $k=0.3$) and one which is overdamped
with $k=2.5$. Estimates using the typical
parametric regimes mentioned, e.g., 
in~\cite{treu} suggest that these regimes
could, in principle, be accessible in the
current system. 

\begin{figure}[H]
\begin{center}
\includegraphics[scale=0.35]{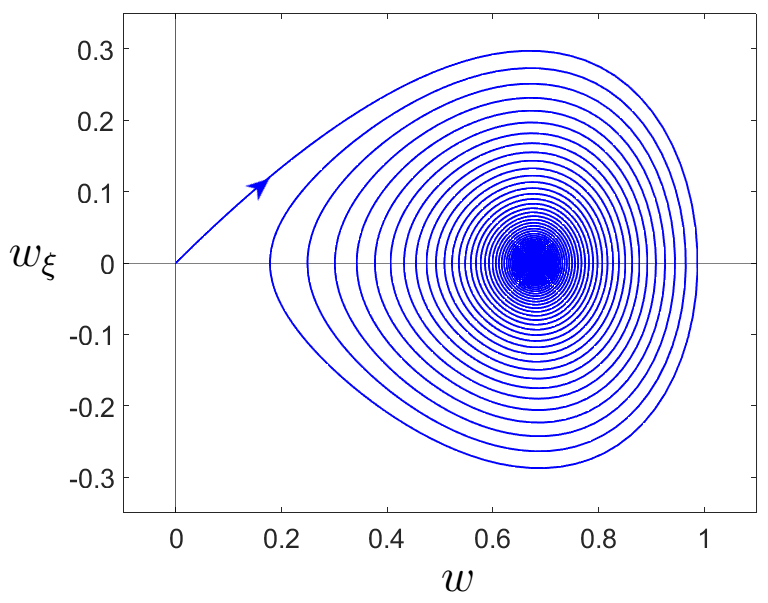}\hspace{0.5cm} \includegraphics[scale=0.35]{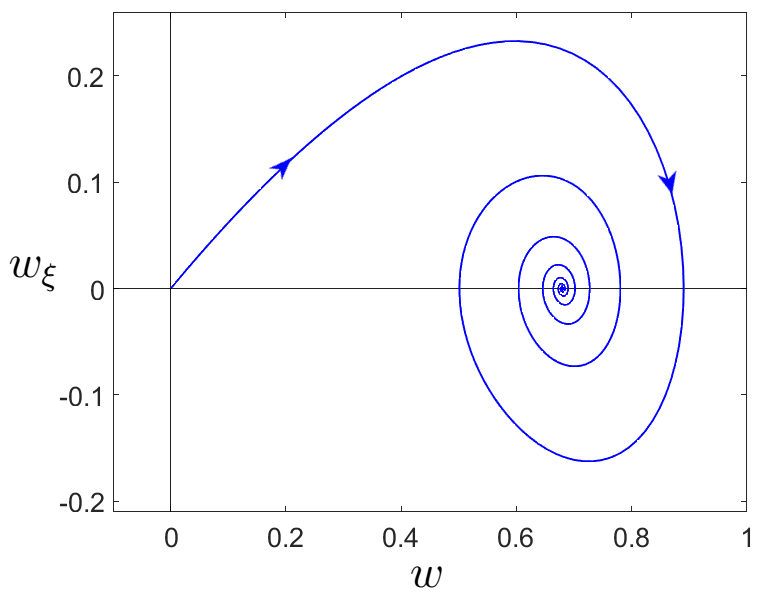}
\caption{When $k>0$ is considered, all physical orbits with $w>0$ end up at the attracting (stable node) equilibrium $w_{min}$. Left panel: For $k=0.03$ an orbit close to the origin is considered which ends up at $w_{min}$ after realizing numerous librations around it. Right panel: The same initial conditions are considered but now for $k=0.3$. The number of librations in this case is significantly decreased.}
\label{fig:dissipative}
\end{center}
\end{figure}

For $k>0$, the homoclinic orbit, as well as the closed orbits of Fig.\ref{fig:conservative} cease to exist. Instead, all orbits (which are limited to positive values of $w$) will end up to the attracting stable node at $w_{min}$. In Fig.~\ref{fig:dissipative} two such cases are shown. In the left panel, the system with $k=0.03$ is examined. We consider an orbit starting infinitesimally close to the origin which is a saddle point. As we can see, the orbit departs from the unstable $w_{max}$ through its unstable manifold and  ends up at the stable equilibrium of $w_{min}$ after orbiting in a spiral way for a long ``time'' (recall that time here represents 
really the effective co-traveling
spatial variable $\xi$) around it. On the other hand, in the right panel of the figure, an orbit with the same initial conditions is considered, but this time in the case of a larger $k$ namely $k=0.3$. The qualitative behavior of the system is the same, but now the orbit performs considerably fewer librations around the equilibrium before it ends up finally at $w_{min}$.

\begin{figure}[H]
\begin{center}
 \includegraphics[scale=0.35]{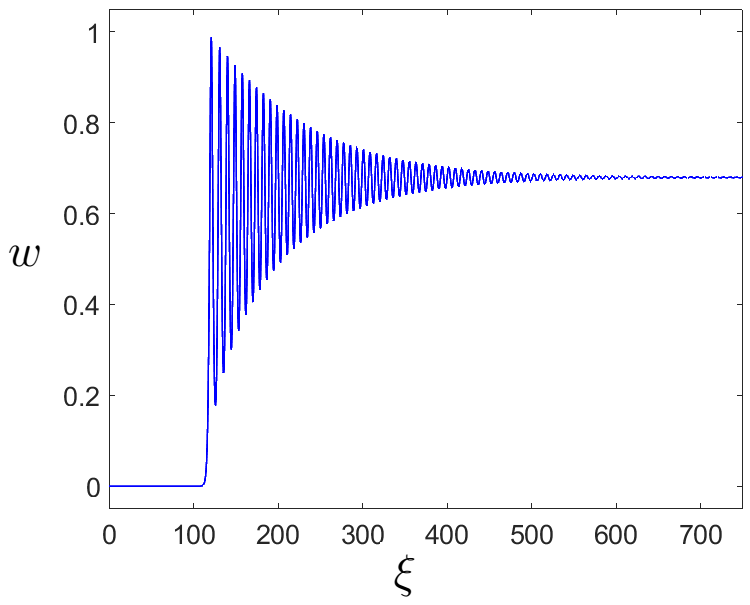}
 \hspace{0.5cm}\includegraphics[scale=0.35]{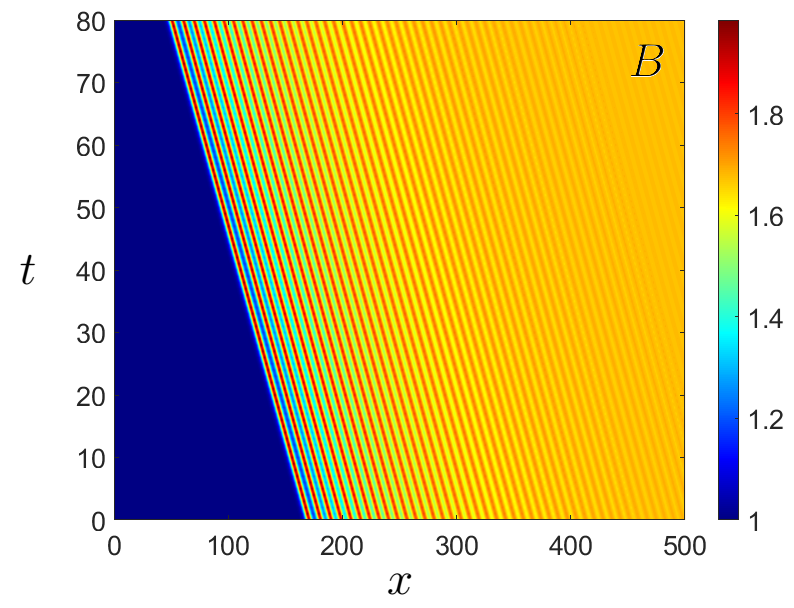} 
\caption{The $k=0.03$ underdamped case. Left panel: the profile of the wave which corresponds to the phase-portrait of the left panel of Fig.~\ref{fig:dissipative}. Right panel: the space-time evolution of the corresponding magnetic field $B$, manifesting the traveling wave
nature of the identified shock wave pattern.}
\label{fig:k_0_03}
\end{center}
\end{figure}

In the $k=0.03$ case, the orbit, in the ``real'' $w-\xi$ space, corresponds to a wave which has the profile of the left panel of Fig.~\ref{fig:k_0_03}. This is a front wave bearing a very long tail. After calculating $w(\xi)$ we can determine $u(\xi)$ by using \eqref{ds12}. Then, we can use the transformation $R=u+R_0$ and $B=w+B_0$ and leverage these initial conditions to numerically integrate the original system \eqref{eq 23}-\eqref{eq 24}. The space-time evolution of this wave for the magnetic field $B$ is shown in the right panel of the figure. The corresponding evolution of the inverse density $R$ is similar to this of $B$ and thus, not shown here.
Notice, despite the complex form of the wave, its genuine (undistorted)
traveling wave form, clearly showcasing for the time scales
considered herein its robust spatio-temporal propagation through
the weakly collisional variant of the Adlam-Allen system.

\begin{figure}[H]
\begin{center}
\includegraphics[scale=0.35]{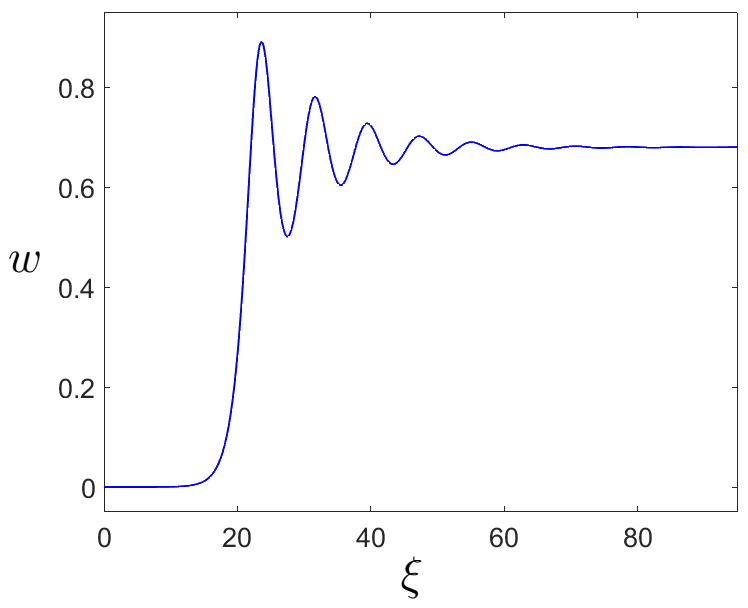}
\hspace{0.5cm}\includegraphics[scale=0.35]{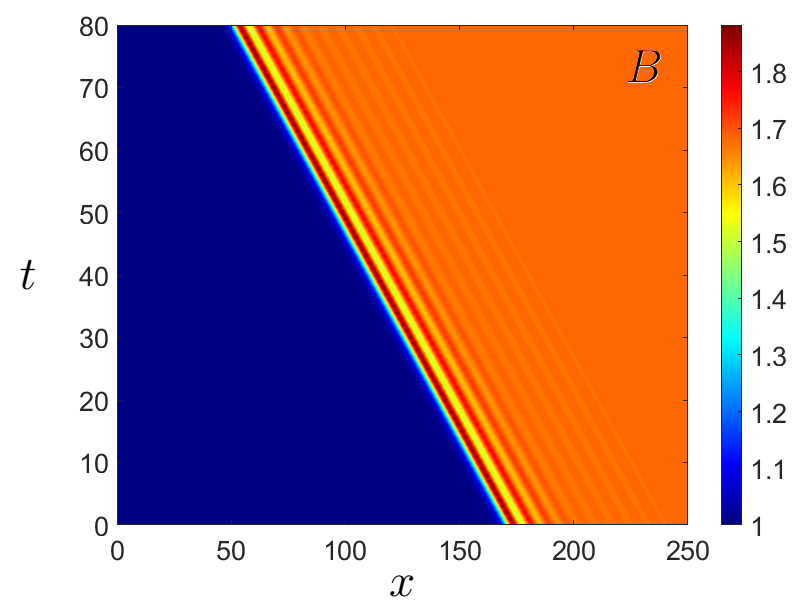}
\caption{The $k=0.3$ underdamped case. Left panel: the profile of the wave which corresponds to the phase-portrait of the right panel of Fig.~\ref{fig:dissipative}. Right panel: the space-time evolution of the corresponding the magnetic field $B$.}
\label{fig:k_0_3}
\end{center}
\end{figure}

In the $k=0.3$ case, the profile of the wave looks like  the one of the left panel of Fig.~\ref{fig:k_0_3}. This is of the same form as the one of Fig.~\ref{fig:conservative} but bears a significantly shorter tail. The corresponding space-time evolution of this wave is shown in the right panel of the figure, manifesting once again the robust
propagation of the structure through the cold weakly collisional plasma
system of interest.

\begin{figure}[H]
\begin{center}
\includegraphics[scale=0.26]{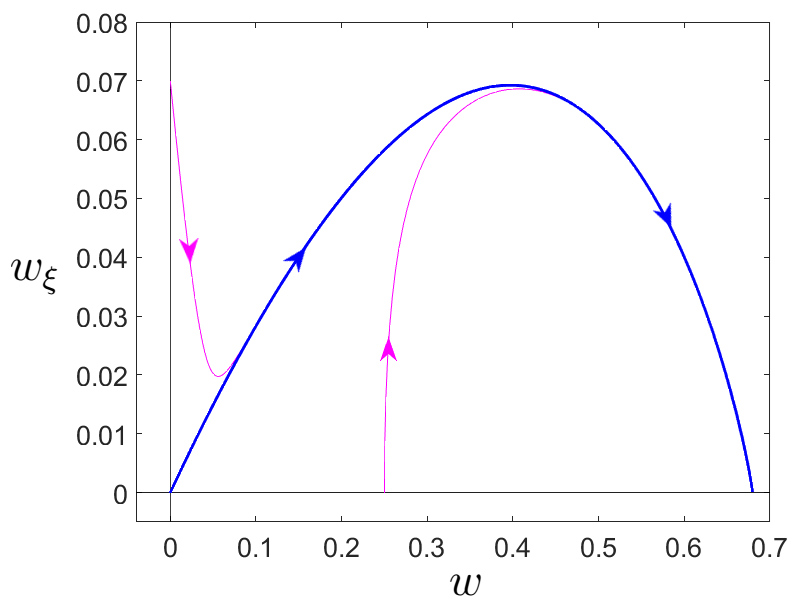}
\includegraphics[scale=0.26]{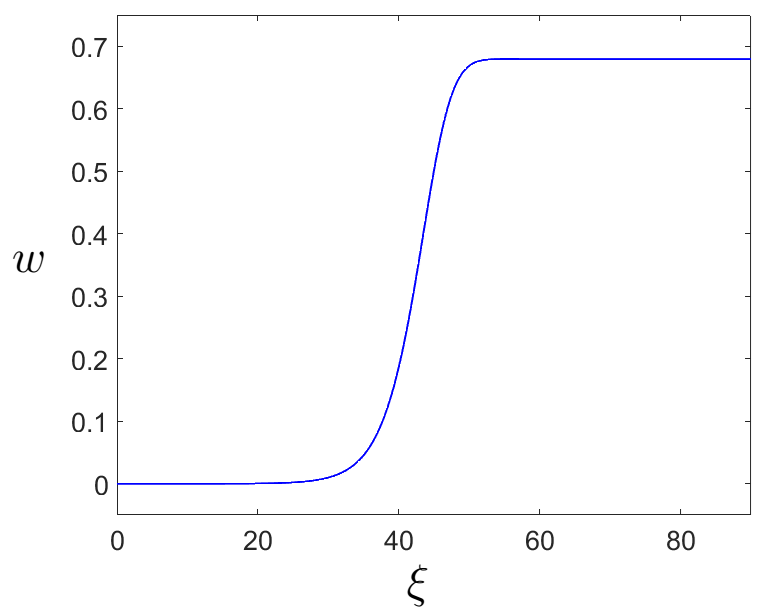}
\includegraphics[scale=0.26]{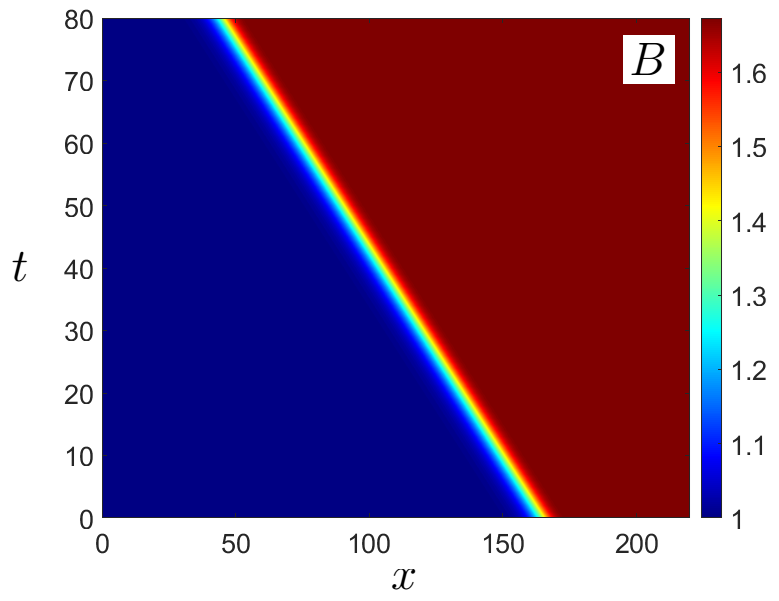}
\caption{The $k=2.5$ (overdamped) case. Left panel: The blue curve corresponds to the heteroclinic orbit between $w_{max}$ and $w_{min}$. The magenta curves depict orbits starting from different points of the phase-plane which quickly converge to the heteroclinic one. Middle panel: The profile of the wave corresponding to the relevant orbit. Right panel: The space-time evolution of the corresponding magnetic field $B$, 
manifesting in this case a robust, monotonic traveling (shock) wave.}
\label{fig:k_2_5}
\end{center}
\end{figure}

We proceed in our investigation by examining the overdamped case of $k=2.5$. 
The results of the study are shown in Fig.~\ref{fig:k_2_5}. In the left panel of the figure, the phase-portrait of the system is depicted. There, we can see that the heteroclinic connection (in blue) between the two equilibrium points is much simpler, i.e., without any 
spiraling around the local minimum. 
This behavior is also supported by the linear stability analysis of $w_{min}$. The eigenvalues of the linearized system (for $B_0=R_0=C=1$) around this equilibrium point are
$$\lambda_{1,2}=\frac{-k\pm\sqrt{k^2-8 c^2-1+3 \sqrt{8 c^2+1}}}{2 c}$$
or, for the considered value of the velocity $c=1.5$,
$$\lambda_{1,2}=\frac{1}{3} \left(-k\pm\sqrt{k^2-19+3 \sqrt{19}}\right).$$
We can see that there is a critical value $k_{cr}$ for which, when $k>k_{cr}$ the complex eigenvalues $\lambda_{1,2}$ become real and negative and thus the stable spiral becomes a stable node. This value of $k$ is $k_{cr}=\sqrt{19-3 \sqrt{19}}\simeq2.43$, which is marginally smaller than the $k=2.5$ value considered. I.e., this case represents the difference
between an underdamped and an overdamped oscillator setting. 
In the same panel, we can also see that orbits originating from different points of the phase-space are also very quickly converging 
towards the homoclinic one, to end up finally (also) at $w_{min}$. In the middle panel of Fig.~\ref{fig:k_2_5}, we can see the 
spatial profile of the left-moving kink in the co-traveling
variable $\xi$, while, in the right panel of the figure the space-time evolution of the corresponding magnetic 
field $B$ is shown. Once again, the relevant
waveform robustly travels undistorted through the medium.

\vspace{-4mm}

\subsection{Right-moving waves - Energy considerations}
As a final aspect of our considerations, we examine a right-moving wave solution of \eqref{eq 25}-\eqref{eq 26}. 
{In this case, we consider a negative value for the velocity $c<0$ which leads to a negative $k/c$ factor for the dynamical system \eqref{ODE}.}

\begin{figure}[H]
\begin{center}
\includegraphics[scale=0.25]{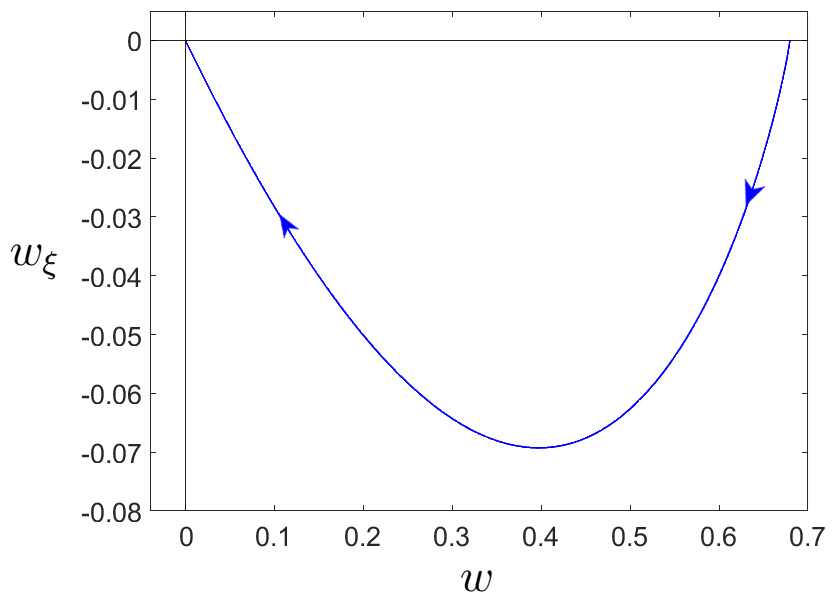}
\includegraphics[scale=0.25]{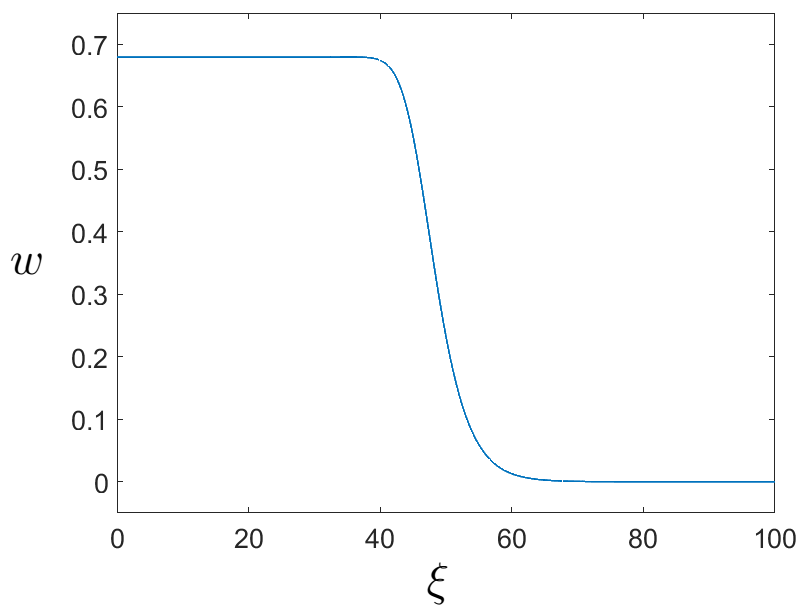}
\includegraphics[scale=0.25]{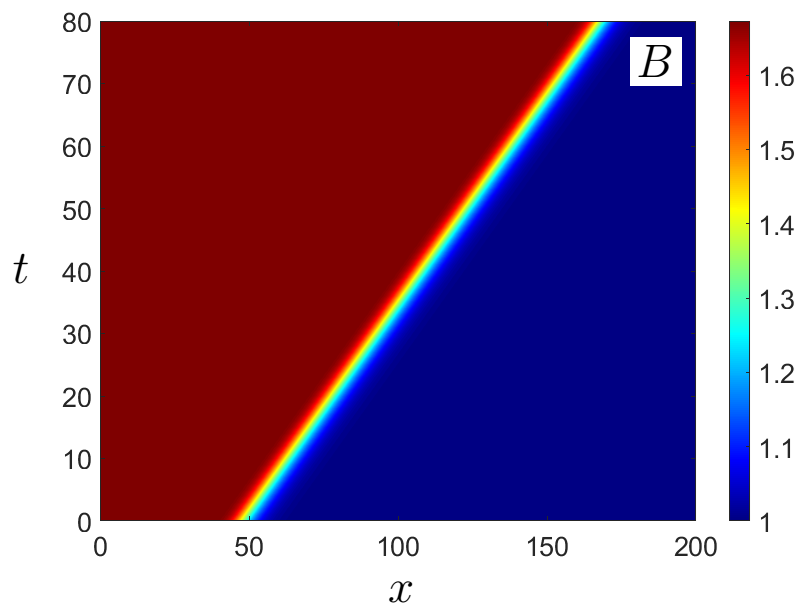}
\caption{The right-moving wave case for $k=2.5$. Left panel: the heteroclinic orbit between $w_{min}$ and $w_{max}$. Middle panel: the profile of the wave corresponding to the heteroclinic orbit.
Right panel: the space-time evolution of the corresponding magnetic field $B$.}
\label{fig:k_2_5_right}
\end{center}
\end{figure}

This case corresponds to a system in which energy is pumped in, i.e.,
an anti-damped setting. In this case, the equilibrium corresponding to the local maximum $w_{max}=0$ remains a saddle-point, while the one corresponding to $w_{min}$ becomes an unstable node. There is still a heteroclinic connection between these two points, but now it starts from the node to end up at $w=0$ through the stable manifold of the saddle-point. We can calculate these heteroclinic orbits for all the values of $k$ considered in the previous section. In order to avoid repetition, we just showcase one of them, namely for the case of $k=2.5$. The results for this choice are shown in Fig.~\ref{fig:k_2_5_right}. In the left panel of the figure we see the heteroclinic orbit from $w_{min}$ to $w_{max}$ passing through negative values of $w_\xi$. In the middle panel, the resulting wave profile is depicted, which resembles a reversed version of the corresponding one of Fig.~\ref{fig:k_2_5}. Finally, in the right panel we can see the space-time evolution for the corresponding magnetic field $B$ observing that now the 
(monotonic) wave is moving to the right.
Notice that we have verified that even a slightly
``inaccurate'' initial condition (e.g., a $tanh$) interpolating between the ``appropriate'' asymptotic states
would still produce the relevant traveling wave.

Interestingly, the energy pumping to the dynamical system 
occurring at the level of our anti-damped oscillator 
enables suitable orbits to escape to infinity. Yet, 
for the traveling waves at the PDE setting considered herein, instead, it just causes the calculated waves to move in the opposite direction. 
{This is also
evident at the level of the relevant ODE of Eq.~(\ref{ODE}) which, by considering as the energy of the system the Hamiltonian of the corresponding conservative system:
\beq
E=H_{c}=\frac{w_{\xi}^2}{2}+V(\xi),
\eeq
leads to the energy balance equation
\beq
\frac{dE}{d\xi}=-\frac{k}{c}w_{\xi}^2,
\eeq
which, in turn, for $c<0$ pumps ``energy'', enabling the orbit
to move
from $w_{min}$ to $w_{max}$.}

\section{Conclusions \& Future Challenges}

In the present work we have revisited a variant of the Adlam-Allen
model, incorporating the dissipative effect introduced in the
context of a cold weakly collisional plasma. In addition to 
reconstructing the relevant effective partial differential equations
at the level of generalized Adlam-Allen model, we have leveraged
the techniques of multiple scale expansions to reduce the relevant
PDEs to a Korteweg-de Vries Burgers equation. The latter prompts
us to expect the existence of shock waves that, depending on the
strength of dissipation, may have an oscillatory or a monotonic 
structure. We have examined both scenarios and identified
the pertinent waves in the AA setting.
Finally, we have explored an ``anti-damped''
scenario where energy may be pumped in the system finding in that
case (perhaps somewhat counter-intuitively) 
that the wave would simply travel in the opposite direction.
Our theoretical predictions based on the multiple scale reduction
analysis and the co-traveling frame dynamical system considerations
are fully corroborated by systematic numerical simulations that
enable us to identify the relevant shock-like patterns and verify
their robust propagation in the full (original) PDE setting.

Naturally, this work paves the way for a considerable array of 
further explorations. For one thing, we have alluded to the
dynamical robustness of the identified traveling waves but have
not considered the spectral or more general stability properties
of the waves we identified. That is a natural step for future studies.
Also, while in the present setting we only considered the
effect of losses (through the collisions incurred, primarily,
by the electrons),  often in similar settings~\cite{Kivshar1989},
it is of interest to examine how the waves are modified in the presence
of a gain source that may balance the lossy character. 
These balanced loss-gain systems may achieve the preservation
of the character of their original solitary waves, under suitable
conditions of gain-loss balance. Finally, to the best of our knowledge,
the vast majority of Adlam-Allen model considerations has been 
restricted to one-dimensional settings. By clarifying the relevant
assumptions and model formulation (and the role of dissipation therein),
we hope that this work may pave the way for the more extensive
consideration of such settings in higher spatial dimensions which may
be of interest in their own right. Such studies are currently 
in progress and will be reported in future publications.

{\it Acknowledgements}. This material is based upon work supported by the U.S. National Science Foundation under the awards PHY-2110030, PHY-2408988, DMS-2204702 (PGK). 
The authors are grateful to Profs. J. Allen, 
D. Frantzeskakis and N. Karachalios 
for discussions on related topics.

\bibliographystyle{unsrt}  
\bibliography{aa_biblio2}       

\end{document}